\documentclass[12pt,a4paper]{article} 

\pdfoutput=1
\usepackage[numbers,sort&compress]{natbib}
\usepackage[affil-it]{authblk}

\usepackage{amsmath}
\usepackage{amsthm}
\usepackage[english]{babel}
\usepackage{graphicx}
\usepackage{amsfonts} 
\usepackage{amssymb} 
\usepackage{hyperref} 
\usepackage{float}
\usepackage{times}
\usepackage{tensor}
\usepackage{mathtools}
\usepackage{color}
\usepackage[font=small,labelfont=small]{caption}
\usepackage{pdflscape}
\usepackage{rotating}
\usepackage{multirow}
\usepackage{array}

\newcommand{\PD}[2]{\frac{\partial #1}{\partial #2}}

\newcommand{\half}[0]{\frac{1}{2}}

\title{An extension of cosmological dynamics with York time} 
\author{Philipp Roser}
\affil{Department of Physics and Astronomy,\\ Clemson University, Kinard Laboratory,\\ Clemson, SC 29631-0978, USA}
\date{\vspace{-0.5cm}}		%ELIMINATES DATE

\begin{document}
 \maketitle
 %\tableofcontents
%------------------------------------
\begin{abstract}
It has been suggested that the York parameter $T$ (effectively the scalar extrinsic curvature of a spatial hypersurface) may play the role of a fundamental time parameter. In a flat, forever expanding cosmology the York parameter remains always negative, taking values $T=-\infty$ at the big bang and approaching some finite non-positive value as $t\rightarrow\infty$, $t$ being the usual cosmological time coordinate. Based on previous results concerning a simple, spatially flat cosmological model with a scalar field, we provide a temporal extension of this model to include `times' $T>0$, an epoch not covered by the cosmological time coordinate $t$, and discuss the dynamics of this `other side' and its significance. We argue that the extension is necessary if a consistent quantisation scheme is to exist. Furthermore, we investigate which types of potentials lead to smooth transitions, paying particular attention to currently favoured inflaton potentials.
\end{abstract}
%------------------------------------

\section{Introduction}
The York time parameter $T$, so called as a result of its first appearance in York's work on the initial value problem (IVP) of general relativity \citep{York1972,York1973,ChoquetBruhatYork1980}, possesses features that make it a compelling candidate for a physically meaningful time. Other than the importance of its role in the IVP, foremost of these are the close relationship between general relativity on constant-mean-curvature (CMC) slices and `shape dynamics' \citep{GomesGrybKoslowski2011, Mercati2014}, and reasons based on the suitability of this parameter choice for quantisation, especially Hidden-Variable approaches \citep{Valentini1996,Valentini1997,Valentini2008inCraigSmith}. Since the extension proposed in this paper hinges on taking the York parameter to be a physically meaningful notion of time it is worth to briefly expand on these motivations.

The initial-value problem of general relativity consists of the task to construct complete initial data on a spatial slice so that the future evolution is fully determined. In general this is difficult since a generic set of spatial metric functions $g_{ij}(x)$ and conjugate momenta $\pi^{ij}(x)$ will not satisfy both the momentum and Hamiltonian constraints. However, on CMC slices it is possible to specify functions $\bar{g}_{ij}(x)$, $\bar{\pi}^{ij}(x)$ which satisfy the momentum constraints only and then rescale the metric, $g_{ij}(x)=\chi^4(x)\bar{g}_{ij}(x)$, by a factor chosen so that $g_{ij}(x)$ and $\pi^{ij}(x)$ satisfy the Hamiltonian constraint, which involves solving an elliptic equation (the so-called Lichnerowicz equation) for $\chi(x)$. Local scale is therefore set apart from the remaining dynamical variables in that it cannot be specified independently. This suggests that this choice of foliation may be physically meaningful and not merely a mathematical convenience.

Shape dynamics is a theory of gravity in which the 4-covariance of general relativity is traded for 3-covariance on spatial slices together with a local symmetry of scale. Roughly speaking, the theory may be considered a natural consequence of the insights gained from York's solution to the IVP. While it shares many physical solutions with general relativity (and could therefore viably describe gravity in our universe), it excludes others (such as closed time-like curves, and differs in its description of black holes, for example \citep{Gomes2014,GomesHerczeg2014}) and has further features that allow it to overcome the problem of time \citep{BarbourKoslowskiMercati2013ProbOfTime}. A particular gauge choice can be made in which shape dynamics becomes locally equivalent with general relativity on CMC slices, although solutions may differ in global structure. 

Reasons for the suitability of CMC slicing for quantisation include the uniqueness of this slicing at least for closed cosmologies \citep{MarsdenTipler1980} and the absence of local singularities which are not part of the final singularity at $T=\infty$ \citep{QadirWheeler1985}. This implies that the necessarily non-local features of a quantum theory can be made well-defined on the metric's configuration space. The non-locality is most explicit in Hidden-Variable approaches such as pilot-wave theory, although even `standard' quantum mechanics (if well-defined in a cosmological setting) cannot escape them. Generic other slicings, such as that in terms of cosmological time (lapse $N=1$), do not possess this feature and contain local singularities on slices of finite $t$, making the construction of a quantum theory difficult.

These developments suggest taking York time seriously in its own right.

In \citep{RoserValentini2014a} we performed a Hamiltonian reduction in order to arrive at the York-time dynamics for a minisuperspace model with a scalar field. Our present purpose is to present an extension of this and similar cosmological models beyond what is usually regarded as temporal infinity. This is done by extending the York time line to the entire real line. We will argue that one ought to take such an extension seriously if York time is indeed to function as a fundamental time parameter and as the time parameter used for quantisation.

In section \ref{background} we will briefly recall the relevant notions relating to the York-time Hamiltonian reduction and discuss some features of the dynamics that we left unaddressed in \citep{RoserValentini2014a} but that are relevant in what is to follow. In the following section (\ref{extension}) we introduce the proposed cosmological extension, based on properties of the York time parameter. Next (sec.\ \ref{mainreason}) we discuss why the extension is required if the theory is to allow for consistent quantisation. We note that scalar fields may display a variety of different behaviours at the transition point where the universe passes from `our side' to the extension. These behaviours are found to depend on properties of the potential and in \ref{potentials} we classify potentials accordingly, paying particular attention to observationally favoured inflaton potentials. Finally, we discuss the relevance of these findings in section \ref{discussion}.

\section{Cosmology with York time}\label{background}

The York parameter $T$ is proportional to the trace of the extrinsic curvature, $K=g^{ij}K_{ij}$, where $g_{ij}$ is the 3-metric on a spatial hypersurface. In particular, $T=(12\pi G)^{-1}K$, where $G$ is the gravitational constant. Slices on which $T$ is constant are therefore slices of constant mean curvature (CMC). For a review of the notion of extrinsic curvature and associated concepts, we refer the reader to \citep{RoserValentini2014a} and references therein, in particular \citep{MisnerThorneWheeler1973}. In a homogeneous cosmological model the York parameter  is furthermore proportional to the Hubble parameter, $T=-\frac{1}{4\pi G}\frac{\dot{a}}{a}$, where $a$ is the cosmological scale factor and the dot denotes differentiation with respect to cosmological time. So the fractional rate of spatial expansion is a measure of York time. For an expanding (contracting) universe, it is the case that $T<0$ ($T>0$). This relationship between the York time and the Hubble parameter is central to the extension we investigate in this paper.

In order to obtain a Hamiltonian description for the cosmological dynamics in terms of York time one may carry out a Hamiltonian reduction \citep{ChoquetBruhatYork1980}. One begins with the ADM decomposition of the Einstein-Hilbert action \citep{ADM1962}, chooses a time parameter $T$ from the extrinsic or intrinsic gravitational variables and then solves the Hamiltonian constraint $\mathcal{H}=0$ for $P_T$, the momentum conjugate to $T$. For the choice of York time these are $T\equiv 2\,tr(\pi^{ij})/3\sqrt{g}$ and $P_T\equiv-\sqrt{g}$, where $\pi^{ij}$ is the momentum conjugate to the spatial metric $g_{ij}$ and $g=\det(g_{ij})$. The reduced (non-vanishing) Hamiltonian which generates the dynamics in terms of York time is then given by $H\equiv-P_T$. 

In general the Hamiltonian constraint is an elliptic equation for $P_T$ and is difficult to solve \citep{ChoquetBruhatYork1980}. However, in the homogeneous isotropic case the equation becomes a depressed cubic, which readily gives an expression for $P_T$ and hence the Hamiltonian. For the case of scalar fields $\phi_A$ in a flat cosmology ($k=0$), which we will consider when constructing our extension and which may, in fact, describe the actual universe (current observation gives $\Omega_k=0.00^{+0.0066}_{-0.0067}$), the Hamiltonian takes the comparitively simple form
\begin{equation} H = \pm \frac{\sqrt{p_\phi^2}}{\sqrt{12\pi GT^2-2V(\phi)}} \qquad\qquad\text{where }p_\phi^2=\sum_Ap_{\phi_A}^2 \label{Hamiltonian}\end{equation}
giving for Hamilton's equations:
\begin{align}
 \phi_A^\prime &= \pm\frac{p_{\phi_A}/\sqrt{p_\phi^2}}{\sqrt{12\pi GT^2-2V(\phi)}} \label{phiprime} \\
 p_{\phi_A}^\prime &= \mp \frac{\sqrt{p_\phi^2}}{(12\pi GT^2-2V(\phi))^\frac32}\PD{V}{\phi_A}. \label{pphiprime}
\end{align}
A prime ($^\prime$) denotes a derivative with respect to $T$. Note that the scale variable ($\sqrt{g}\sim a^3$) and its conjugate momentum ($tr(\pi^{ik})$) have been eliminated in favour of explicit time dependence. This is a general feature of non-trivial Hamiltonian reductions. 

In \citep{RoserValentini2014a} we discussed the classical and quantum theory for the case of a single scalar field. Here the correct Hamiltonian reads $H=\pm |p_\phi|/\sqrt{12\pi GT^2-2V(\phi)}$. In a minor oversight there we wrote $p_\phi$ rather than $|p_\phi|$ in the numerator.

The sign ambiguity arises from the fact that both sign choices for $H$ solve the Hamiltonian constraint. The choice is arbitrary. The set of solutions for $(\phi_A(T),p_{\phi_A}(T))$ is identical with the exception of opposite signs in the corresponding functions $p_{\phi_A}(T)$. The physical interpretation of the numerical value of $H$ is `volume' (since $H\equiv-P_T=\sqrt{g}\propto a^3$, $a$ being the scale factor), from which one may be inclined to infer that $H>0$. However, in general the scale factor is defined as part of the metric, that is, as defining lengths $dl^2=a^2\delta_{ij}dx^idx^j$, which due to its `square' nature strictly speaking leaves the sign of $a$ (and hence $a^3$) ambiguous and unphysical. Here we choose the natural convention that $a>0$ and therefore the positive sign in the expression for $H$.

By using the lapse, found to be $N_T=-(4\pi GT)^{-1}\frac{a^\prime}{a}$, it is relatively straightforward to show that these equations combined with the interpretation of the value of the Hamiltonian $H=a^3$ are equivalent to the standard Friedmann-Lema\^{i}tre equations for a set of scalar fields in a flat cosmology,
\begin{align} 
 \frac{\dot{a}^2}{a^2} &= \frac{8\pi G}{3}\rho \label{eq:FL1}\\
 \frac{\ddot{a}}{a}    &= -\frac{4\pi G}{3}(\rho+3P), \label{eq:FL2}
\end{align}
where $\rho=\sum_A\frac12\dot{\phi}_A^2+V(\phi)$ and $P=\sum_A\frac12\dot{\phi}_A^2-V(\phi)$, together with the Klein-Gordon equations
\begin{equation} 
  0 = \ddot{\phi}_A+3\dot{\phi}_A\frac{\dot{a}}{a}+\PD{V}{\phi_A}. \label{eq:KG}
\end{equation}
where $\dot{\phi}_A=d\phi_A/dt$ and $t$ is the conventional cosmological time parameter.

A relevant feature of this dynamics which we did not discuss in \citep{RoserValentini2014a} (in part due to the oversight mentioned above) is that of `turning points', points where $\phi_A^\prime=0$ (or equivalently $\dot{\phi}_A=0$) for all $A$ simultaneously. In general such points are not generic (only a measure-zero subset of solutions has them) except for the scenario where only a single scalar field is considered, so let us focus on the latter. One finds that at such points $12\pi GT^2-2V(\phi)=0$ (one may easily infer from eq.\ \ref{eq:FL1} and $T=-\frac{1}{4\pi G}\frac{\dot{a}}{a}$ that $\dot{\phi}^2\propto12\pi GT^2-2V(\phi)$, which also guarantees the non-negativity of the radicand). It appears that the expressions \ref{Hamiltonian}, \ref{phiprime} and \ref{pphiprime} become singular.

Let us frame our discussion entirely within the language of the York-time dynamics without reference to cosmological time $t$. After all, it is our goal in this paper to consider York time as physically preferred. A turning point is defined by $\phi^\prime=0$, so by eq.\ \ref{phiprime} whose numerator simply reads $sign(p_\phi)$ in the case of a single scalar field we must have $p_\phi=0$. But then either $H=0$ (and therefore $a^3=0$, constituting an initial or final singularity), or ---if the turning point is to occur at finite volume--- $12\pi GT^2-2V(\phi)\rightarrow0$ as the turning point is approached, specifically $(12\pi GT^2-2V(\phi))^\half\sim p_\phi$. The implication is that $\phi^\prime\rightarrow0$ with opposite signs for approach of the turning point from the future and past. %DISCUSS LIMIT OF DELTA-$\phi$ INTEGRATING OVER SMALL INTERVAL AROUND TURNING POINT. GUARANTEED EQUIVALENCE FOR TRAJECTORIES

\begin{figure}[ht]
 \includegraphics[width=0.4\textwidth]{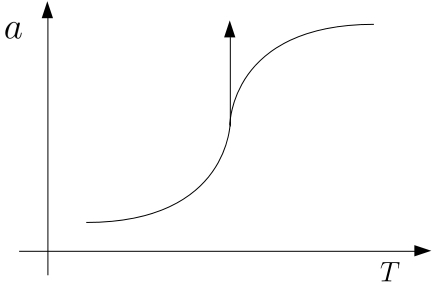}
 \caption{Schematic illustration of the evolution of the scale factor near a turning point. The rate of growth becomes infinite at a single point but there is no discontinuity in the scale factor $a$.}
 \label{turningpointgraphic}
\end{figure}

In order to understand the nature of these turning points, consider the `interpretation' of the Hamiltonian as volume, $H\propto a^3$. Using expression \ref{Hamiltonian} and differentiating with respect to $T$ gives
\begin{equation} \frac{a^\prime}{a} =-\frac{4\pi GT}{12\pi GT^2-2V(\phi)}, \end{equation}
so if the turning point is to occur at finite volume (and therefore finite $T$), we see that $a^\prime/a\rightarrow\infty$ as we approach the turning point. At the turning point the fractional rate of expansion (with respect to York time) diverges, although the total fractional growth remains finite and there is no discontinuity in $a$ (see fig.\ \ref{turningpointgraphic}). This corresponds to a de~Sitter phase (of infinitesimal duration in terms of cosmological time), where the fractional rate of expansion is constant and the York lapse diverges, $N_T\rightarrow\infty$.

Note that even in a universe with only a single scalar field such turning points do not necessarily occur, but their existence depends on the form of the potential $V(\phi)$, with consequences for the late-time ($T\rightarrow0$)\footnote{Recall that we are discussing the case of a flat cosmology, which is forever expanding (when phrased in terms of cosmological time). There are models where a universe with a flat cosmology recontracts within finite cosmological time (\citep{GiamboMiritzisTzanni2014} and references therein), but these do not fall within the scope of this paper since an extension of York time does not exist in these cases as the full the real line of York time is already in one-one correspondence with the history of the universe.} behaviour of $\phi$. For example, in the absence of turning points and $V(\phi)>0$ everwhere it is the case that $\phi$ necessarily diverges at least logarithmically in the late-time limit. This follows from the non-negativity of $12\pi GT^2-2V(\phi)$, which implies that $\phi^\prime\geq(12\pi GT^2)^{-\frac12}$.

We will return to the dependence of the late-time behaviour on the form of the potential and specifically the existence of turning points in section \ref{potentials}.

\section{Extending the cosmology beyond \texorpdfstring{$T=0$}{T=0}}\label{extension}

Before constructing the proposed extension, let us briefly examine the case of a closed ($k>0$) rather than flat ($k=0$) cosmology. Here the fractional rate of expansion approaches infinity as we go back in time towards the initial singularity, reaches zero after some finite time when the universe has its maximal size and then recollapses, approaching an infinite negative fractional rate of expansion. Hence the York parameter is $-\infty$ at the initial singularity, zero at the time of maximal volume and approaches $+\infty$ near the final singularity. Furthermore, the York parameter is monotonic. Therefore $T$ seen as a function from points along the history of the universe to the real numbers (the answer to `what York time is it?') is a bijection. 

In a flat cosmology the situation is different. Instead of reaching a point of finite maximal volume the universe continues to grow, either with the Hubble parameter approaching zero (if $V(\phi(t))\rightarrow0$ as $t\rightarrow\infty$, or approaching a constant non-zero value if $V(\phi)$ includes a positive cosmological constant or is otherwise such that $V(\phi(t))\rightarrow V_\infty>0$. Then $T\rightarrow T_\infty$ as $t\rightarrow\infty$ with $T_\infty=0$ in the former case and $T_\infty<0$ in the latter. York time `ends' at zero or a finite negative value. We now consider extending the York time line beyond $T_\infty$, assuming the dynamical equations \ref{phiprime} and \ref{pphiprime} to continue to hold. The idea is that if York time is supposed to be in some sense \emph{the} physically fundamental time parameter, then a piori there is no reason why $T$ should end at some particular or indeed any finite value. In fact, as we discuss in the next section there are physically motivated reasons why such an extension may be necessary. In the language of cosmological time the end is infinitely far away, but in terms of York time it is nigh indeed: Measurements of the Hubble parameter place the current value of York time at around $6.2\cdot10^{-61}$, so that $T\sim10^{-60}$ in reduced Planck units.

A reader puzzled by this idea may find it helpful to consider an analogy. In the case of a Schwarzschild black hole a new region of space-time is revealed by the use of Kruskal-Szekeres coordinates, extending to the `other side' of the singularity (where the Kruskal-Szekeres variable usually denoted by $V$ takes negative values), a region not described by either Schwarzschild or Eddington-Finkelstein coordinates \citep[Ch.\ 31]{MisnerThorneWheeler1973}.\footnote{Arguably there is another, simpler example: the extension of dynamics beyond a black hole's event horizon in the first place. The idea that spacetime extended beyond the horizon of a Schwarzschild black hole into the region $r<2M$ took several decades to become apparent (\citep{Eddington1924,Lemaitre1933,Synge1950}). However, a crucial difference here is that this extension concerns the removal of a mere coordinate singularity. No physical quantities become infinite (and geodesics crossing the surface $r=2M$ are traversed in finite proper time).} 

At least two questions require answering. First, how do the dynamical entities ---in our case the scalar field--- evolve during the transition? Second, what evolution can we expect on the other side?

Let us answer the second question first. Equations \ref{phiprime} and \ref{pphiprime} are invariant under temporal reflection about $T=0$ (with the direction of time unchanged, so that $dT$ does not transform). Furthermore, $sign(\phi^\prime)=sign(p_\phi)$. Therefore the set of solutions for the interval $T\in(0,\infty)$ is exactly the same set of trajectories as for the interval $T\in(-\infty,0)$ except that the latter are traced out `backwards' and the sign of the momentum is flipped. A solution $(\phi(T),p_\phi(T))$ for $T<0$ may be used to define a corresponding solution $\bar{\phi}(T), \bar{p}_\phi(T)$ for $T>0$ via initial data at $T_0>0$ given by 
\begin{equation}\bar{\phi}(T_0)=\phi(-T_0),\quad\bar{p}_\phi(T)=-p_\phi(-T),\end{equation} 
as can be easily verified. 

The equivalence between equations \ref{phiprime} and \ref{pphiprime} together with $H=a^3$, and the conventional Friedmann-Lema\^itre and Klein-Gordon equations (\ref{eq:FL1}, \ref{eq:FL2}, \ref{eq:KG}) implies that we should also be able to understand this symmetry in terms of the latter, even though the cosmological time parameter $t$ does not cover the period $T>0$. Instead we are able to define another analogous cosmological time parameter $t^\prime$ such that $t^\prime\rightarrow-\infty$ as $T\rightarrow0_+$ and $t^\prime\rightarrow0$ (or some other finite value) as $T\rightarrow\infty$. This parameter $t^\prime$ does not describe the period $T<0$. Equations \ref{eq:FL1}, \ref{eq:FL2}, \ref{eq:KG} are invariant under reversal of the direction of time, $dt\rightarrow-dt$. One can therefore see that if $t$ is mapped to $-t^\prime$ as part of the correspondence between solutions then these equations remain satisfied for the time variable $t^\prime$ instead of $t$.

However, note that in general for more complicated matter content there is no reason to expect that the universe will simply `turn around' and trace out its trajectory backwards, even if there is a one-one correspondence between solutions on `our side' and the extension.

Now turn to the transition. Recall that $T$ is proportional to the Hubble parameter, so that $T\rightarrow0$ as $t\rightarrow\infty$ only if $\frac{\dot{a}}{a}\rightarrow0$, which is the case in the presence of a free scalar field or a potential which goes to zero sufficiently fast. Specifically, note that $12\pi GT^2-2V>0$ as we saw above. With $T\rightarrow0$, this implies that if $V\geq0$ everwhere, then $V\rightarrow0$ faster than $6\pi GT^2$ if $T$ is to go to zero. Otherwise, in the presence of a positive cosmological constant, for example, $T$ does not reach zero in finite cosmological time. While in that case there are still corresponding solutions on the `other side', there is an intermediate interval $T\in(T_\infty,-T_\infty)$ where the dynamics is not defined. We will discuss how the nature of the transition relates to properties of the potential in more detail in section \ref{potentials}.

\section{The necessity of the extension}\label{mainreason}

Why should one take the extension seriously? Answering that is the purpose of this section. There may be some purely philosophical arguments: that time should not end at some finite value, or that the mere mathematical possibility of the extension should at least prompt us to allow for the possibility of its physicality. However, there is a considerably more compelling, \emph{physical} reason to consider it more than mere fiction.

The model employed in this paper is that of a homogeneous isotropic universe. This is, of course, an approximation. Let us consider the existence of inhomogeneities in this universe, in particular local singularities such as a Schwarzschild black hole formed at some finite time from a collapsing cloud of dust (see \cite[sec.\ 32]{MisnerThorneWheeler1973} and references therein for an introduction). Such a scenario can be described by smoothly `gluing' together parts of a closed collapsing Friedmann universe shrinking to zero size (the collapsing dust cloud), a Schwarzschild region of space-time (the near exterior of the cloud) and the chosen Friedmann cosmology (the distant rest of the universe). When described in terms of CMC surfaces it is found \citep{BrillCavalloIsenberg1980} that these surfaces smoothly connect between the different regions.\footnote{Interestingly, a description in terms of a conventional time parameter $t$ allows only for continuous surfaces, but not smooth ones.} Furthermore, $K\rightarrow\infty$ ($T\rightarrow\infty$) as one apporaches the singularity formed by the fully collapsed dust cloud, independently of the choice of exterior cosmology.

If the cosmology is closed, all CMC slices inside the cloud and Schwarzschild regions connect smoothly to exterior cosmological slices \citep{QadirWheeler1985}. If the cosmology is flat, only slices up to $K=0$ ($T=0$) do so and slices with $K>0$ only exist in disconnected patches around the local singularities \citep[secs.\ VI, VII]{BrillCavalloIsenberg1980}. An observer falling into the black hole reaches the singularity in finite proper time and therefore crosses all CMC surfaces up to $T=\infty$ in finite proper time. If York time has fundamental physical significance, reaching the singularity constitutes arriving at the end of the universe at $T\rightarrow\infty$ no matter the exterior cosmology. Therefore, without the extension, different areas of the universe would end at different times: at $T=0$ far away from the singularity, but at $T=\infty$ in its vicinity.

Classically this does not immediately cause a contradiction. However, quantisation of the theory is now a problem. The quantum theory is necessarily non-local --- this was one reason motivating the CMC slicing in the first place --- and non-local effects are incompatible with partial spatial slices of this kind. 

For example, consider a pair of entangled particles, one of which enters the Schwarzschild radius while the other remains in some distant part of the universe. When the falling particle reaches $T=0$ its partner encounters the end of time. What happens to the quantum state at this point is entirely unclear. At $T=0$ the configuration space (or Hilbert space) would suddenly change its dimensionality since the space on which the dynamical entities such as matter fields are defined shrinks to a set of disconnected finite patches. This does not allow for a consistent definition of the quantum theory. Furthermore, since the falling particle crosses into the region $T>0$ in finite proper time this scenario cannot be dismissed as unphysical.

Consistent quantisation therefore demands the existence of the extension.

\section{Classification of potentials by behaviour at the transition}\label{potentials}

\begin{figure}[ht]
 \includegraphics[width=\textwidth]{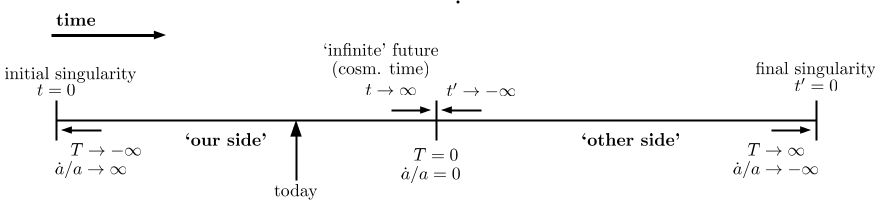}
 \caption{Overview of corresponding values of cosmological time $t$ and York time $T$ in the case of a free scalar field. Only the left half of the time line is described by conventional cosmological time for a flat universe. The right hand side is however a natural extension if the York parameter is taken seriously as a fundamental time parameter. The finite `starting' and `end' values of $t$ and $t^\prime$ respectively are up to convention (due to cosmological-time translation invariance).}
  \label{fig:free}
\end{figure}

We are interested in identifying the class of potentials for which there is a smooth transition from the $T<0$ to the $T>0$ region. Specifically, what we mean by this is that 
\begin{itemize}
 \item the dynamical equations \ref{phiprime}, \ref{pphiprime} describe the behaviour of the scalar field(s) for the entire real line and there is no intermediate interval around $T=0$ where $12\pi GT^2<2V(\phi)$,
 \item the energy density associated with the scalar field and the field value $\phi$ itself remain finite everywhere. %TOTAL ENERGY?
\end{itemize}
The first condition is equivalent to the question in the cosmological-time picture whether or not there are solutions to eq.\ \ref{eq:FL1} for all values of $\dot{a}/a$. For example, the dynamics of the free field ($V(\phi)=0$) can be analytically solved,\footnote{The relation $\phi_{free}^\prime\sim|T|^{-1}$ illustrates a feature of the York time picture: More and more `happens' in a finite time interval the closer we are to $T=0$. It is interesting that a similar observation was made about \emph{cosmological time} regarding the very \emph{early} history (close to $t=0$) in the 1960s by \citet{Misner1969c}, who suggested that $-\ln a$ (or alternatively the logarithm of the homogeneous temperature, which turns out to roughly equivalent) might be a more appropriate choice of temporal parameter to describe the history of the universe. This is further exemplified in our language to describe the early universe in terms of `epochs' such as the `Planck', `Grand Unifying' and `Inflationary' epochs, each of which is described by a vastly different order of magnitude of duration. As Misner states, ``[t]he universe is meaningfully infinitely old because infinitely many things have happened since the beginning'' \citep[p.\ 1331]{Misner1969c}. Another comparable choice is the parameter $\ln t$ as had been advocated by Milne another twenty years earlier (cited in \citep{Misner1969c}).}
\begin{align} 
 \phi_{free}^\prime &= sign(p_\phi^{free})\cdot\big(12\pi GT^2\big)^{-\frac12},\qquad\qquad p_\phi^{free} = \text{const.} \notag\\
		    \Rightarrow\quad\phi_{free} &= (12\pi G)^{-\frac12}\ln|T|
\end{align}
for all $T<0$ (fig.\ \ref{fig:free}). The energy density is $\sim\dot{\phi}^2\sim T^2\rightarrow0$ and therefore obviously finite. However, $\phi$ diverges logarithmically with its sign equal to the sign of the initial value $p_\phi^{free}$. %TOTAL ENERGY?

\begin{figure}[ht]
 \includegraphics[width=\textwidth]{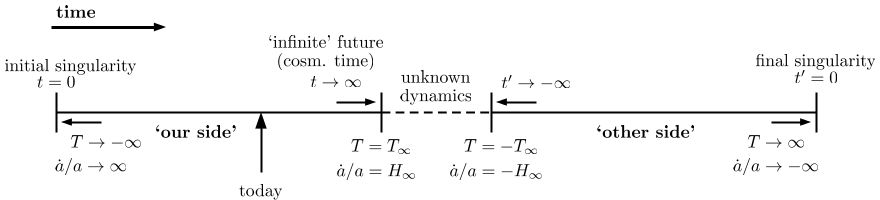}
 \caption{Overview of corresponding values of cosmological time $t$ and York time $T$ in the case of a scalar field with a positive cosmological constant. The left section of the line is described by conventional cosmological time $t$, the right one by a similar parameter $t^\prime$. In the middle section, $T_\infty<T<-T_\infty$, the dynamics of the scalar field is undefined, raising philosophical questions.}
 \label{fig:cosm-const}
\end{figure}

On the other hand, consider the simple case of a positive cosmological constant, $V(\phi)=\Lambda/4\pi G$, $\Lambda>0$, and no other potential term. In terms of cosmological time, this is associated with an eternally expanding universe with $\dot{a}/a$ asymptotically approaching a finite value proportional to $\Lambda$ (de~Sitter expansion). In the York-time picture, $12\pi GT^2-2V(\phi)>0$ only up until $T_\infty=-\Lambda/(48\pi G)$ and again after $-T_\infty=+\Lambda/(48\pi G)$. In the intermediate period the dynamics is ill-defined since $\phi$ was taken to be a real field (fig.\ \ref{fig:cosm-const}). While the dynamical equations can indeed be applied to $T>-T_\infty$, the transition is not smooth. We will provide a discussion of the meaning and implication of such an intermediate period below.

A non-trivial example of a potential which leads to a well-defined transition, but with diverging field value, is given by $V(\phi)=V_0e^{-\lambda\phi/M_P}$, where $V_0$ and $\lambda$ are constant parameters and $M_P=(8\pi G)^{-\frac12}$ is the Planck mass. Potentials of this form may arise in the effective four-dimensional dynamics induced by Kaluza-Klein theories, for example. A discussion of this potential in terms of conventional cosmological time $t$ is given in \citep{FerreiraJoyce1998} and references therein, in particular \citep{Halliwell1987}. This model can be solved exactly. Make the ansatz $\phi(T)=\phi_0+\alpha\ln|T|$, then realising that $dV/d\phi$ is monotonic one can infer that there is a time after which there is no further turning point, so that $sign(p_\phi)$ remains fixed and equation \ref{phiprime} effectively decouples from the momentum. Solving the equation with the proposed ansatz one obtains the solution
\begin{equation}\label{exponentialpotentialsolution}
 \phi(T) = \frac{M_P}{\lambda}\ln\left(\frac{8V_0M_P^2}{(6-\lambda^2)}\frac{1}{T^2}\right),
\end{equation}
which diverges as $T\rightarrow0$. However, 
\begin{equation}\label{exponentialpotentialpotential}
 V(T) = \frac{(6-\lambda^2)T^2}{8V_0M_P^2}
\end{equation}
approaches zero in this limit, as required for the existence of a well-defined transition. Note that due to the fact that $T$ is physically meaningful there is no time-translation invariance (see \citep{RoserValentini2014a}).

A simple example of a fully smooth transition (with $\phi$ remaining finite) is provided by a scalar field with a positive mass term, $V(\phi)=m^2\phi^2$, $m^2>0$ (which is a version of large-field inflation \citep{MartinEtAl2013_EncyclopaediaInflationaris}, albeit one not favoured by the most recent data \citep{Planck2015_Overview}). One does not need to solve the equations in order to analyse the behaviour as $T\rightarrow0$. In the late-time limit one expects oscillatory behaviour with an amplitude decreasing over time. Indeed, $\phi$ begins by rolling down the potential (if its initial conditions are such that it rolls up, then eq.\ \ref{pphiprime} ensures that $p_\phi$ will pass through zero at some point, a turning point, and $\phi$ changes direction) and passes through the minimum, at which point $p_\phi^\prime$ changes sign (eq.\ \ref{pphiprime}). The momentum $p_\phi$ will therefore pass through zero eventually, so that a turning point results and $\phi$ reverses sign. This repeats, resulting in oscillatory behaviour. The amplitude of the oscillations decreases since $12\pi GT^2$ monotonically decreases, so the turning point (recall that these can be identified by the condition $12\pi GT^2-2V(\phi)=0$) are reached at a lower value of $V(\phi)$ and therefore at a lower value of $\phi^2$ than during the previous cycle. It is noteworthy that in the cosmological-time picture appeal to `Hubble friction' must be made (the Hubble parameter appears as a frictional coefficient in eq.\ \ref{eq:KG}), whereas in the York-time equations derived via the Hamiltonian reduction the decreasing amplitude is encoded in the explicit time dependence.

During the oscillations the field $\phi$ remains finite since it is bound by the field value at the most recent turning point from above and below. Both potential and kinetic energy density furthermore approach zero, so the finiteness of the energy density $\rho_\phi$ is guaranteed. In fact, we know that for oscillatory solutions that $\rho_\phi\sim a^{-3}$ (see \citep{FerreiraJoyce1998}), so even the total energy remains finite if initially normalised.

Let us now attempt a more general characterisation. We restrict our discussion to potentials bounded from below. For consider a potential without a lower bound. If there is a local minimum the field may, initial conditions permitting, become trapped in the well and the potential may be locally approximated by another potential bounded from below. If on the other hand no such minimum exists or the initial conditions are such that the field does not become trapped inside one, then $V(\phi)$ reaches arbitrarily negative values, so that $12\pi GT^2-2V(\phi)>\epsilon$ as $T\rightarrow0$ for some finite $\epsilon>0$. Therefore $|\phi^\prime|$ is bounded for all $T$ greater than some reference time $T_r$. Since the interval $(T_r,0)$ is finite, $\phi$ remains finite also. The `transition' is perfectly smooth. However, it is also not a novelty since it is reached in finite cosmological time \citep{GiamboMiritzisTzanni2014}. That is, the universe recontracts even in the conventional description. No new temporal region is revealed through the use of York time.

Therefore, assume there exists $\alpha\in\mathbb{R}$ such that $\alpha=\inf V(\phi)$. Table \ref{TofB} summarises the implications for the transition for different cases.

\begin{landscape}
\begin{figure}[H] 
\begin{center}
\begin{tabular}{|p{0.12\linewidth}|m{0.10\linewidth}|m{0.25\linewidth}|m{0.25\linewidth}|m{0.25\linewidth}|}\hline
   \multicolumn{1}{|c|}{\multirow{2}{*}{\textbf{Value of $\inf V(\phi)$}}} & \multicolumn{1}{|c|}{\multirow{2}{*}{\textbf{Examples}}} & \multicolumn{1}{|c|}{\textbf{Value of $T_\infty$}} 
	  & \multicolumn{2}{|c|}{\textbf{Smoothness of transition}} \\\cline{4-5} 
    & & \multicolumn{1}{|c|}{\textbf{(Existence of Transition)}} & \multicolumn{1}{|c|}{\textbf{Field value finite? ($|\phi|<\infty$)}} 
	  & \multicolumn{1}{|c|}{\textbf{Energy density finite?}} \\\hline 
   $\inf V(\phi) < 0$ 
	& As below but with\newline cosm.\ const.\newline ($\Lambda<0$)
	& \textbf{None.} $T_\infty$ does not exist, the universe recontracts in finite cosmological time. No extension can be made. The null energy condition is violated.
	& \begin{center}\textbf{No real transition.}\newline (Extension does not exist.)\end{center}
	& \begin{center}\textbf{No real transition.}\newline (Extension does not exist.)\end{center} \\ \hline
   $\inf V(\phi) = 0$ 
	& AI, HI,\newline MHI, RGI,\newline SBI
	& $\mathbf{T_\infty=0}$ \textbf{(transition exists)}, unless there is a local minimum with $V(\phi_{min})>0$ in which the field is trapped, in which case behaviour as if $\inf V(\phi)>0$.
	& \textbf{Possibly:}  Yes iff there is an infinite number of turning points (i.e.\ oscillating solutions, requires existence of a minimum) (e.g.\ HI, MHI, RGI, SBI). Otherwise $\phi\rightarrow\pm\infty$ at least logarithmically (e.g.\ AI).
	& \textbf{Yes.}\newline Example: oscillating solutions in $V\sim|\phi|^n$: $\rho_\phi\sim a^{-m}$, $m=6n/(n+2)$ \citep{FerreiraJoyce1998}  \\\hline
   $\inf V(\phi) > 0$ 
	& As above but with\newline cosm.\ const.\newline ($\Lambda>0$)
	& $\mathbf{T_\infty<0}$ \textbf{(intermediate epoch of ill-defined dynamics)}. C.f.\ model with a cosmological constant.
	& \begin{center}\textbf{No transition.}\end{center} (Epoch of ill-defined dynamics before other side reached.)
	& \begin{center}\textbf{No transition.}\end{center} (Epoch of ill-defined dynamics before other side reached.)\\\hline
 
\end{tabular}
\caption{\textbf{Transition behaviour for potentials bounded from below.} The second entry in each line gives examples of potentials with the infimum as given in the first column from the most favoured inflation potentials based on the 2013 Planck data as identified in \citep{MartinEtAl2013} (although here we do not differentiate between subcategories of a particular type of potential based on parameter-value ranges). The third column identifies if York time reaches $T=0$, the condition for a transition to the other side to exist without an intermediate epoch of ill-defined dynamics. The last two columns provide information on the smoothness of the transition.}
\label{TofB}
\end{center}
\end{figure}
\end{landscape}

\section{Discussion}\label{discussion}
The matter content of the real universe is considerably more complicated than a single scalar field. However, during inflation a scalar field may have constituted the dominant constituent of the matter content. Furthermore, with all other contributions to the energy density falling to zero in the late-time limit of the universe, a non-zero contribution from a scalar field potential may become once again significant. Therefore considering a scalar-field model in order to explore the `other side' is appropriate.

The question of the significance of the transition properties of such potentials arises. Indeed, one may dismiss the cosmological extension presented here as purely mathematical speculation without any physical significance at all. However, if one considers the possibility that York time may indeed play the role of a fundamental time parameter that forms the basis of quantisation (where, as we showed, the extension is required for consistency), then the existence of models (potentials) with an intermediate period around $T=0$ of ill-defined dynamics requires consideration. One may take one of the following stances:\footnote{Not including more radical attitudes such as the view that during the intermediate epoch the dynamical evolution of matter and space cannot be derived from a variational principle, or that our mathematical tools are in some way insufficient to provide a description of the dynamics at these times. Whether or not such a view is viable is a philosophical question outside the scope of this paper.}
\begin{enumerate}
 \item Potentials which lead to such an intermediate period are ruled out by a theory in which York time is considered fundamental. Conversely, if observational evidence suggests that inflation, for example, is driven by such a potential, then this would falsify a theory in which York time is taken to play a physically fundamental role.
 \item Time may end at some finite value, so there is no contradiction in regarding York time as physically meaningful but without taking the extension seriously. The argument of section \ref{mainreason} is avoided by postulating some different, non-equivalent and at this stage unknown quantisation procedure which allows time to end at different values in different regions of space, or a change in the classical theory that avoids the `patchy' time slices for $T>0$.
\end{enumerate}
The second point of view may be found viable for example in the context of shape dynamics \citep{GomesGrybKoslowski2011, BarbourKoslowskiMercati2013ProbOfTime,Mercati2014}, where what is presumed physical is a sequence of configurations (that is, a trajectory in configuration space) without any physically fundamental time. A preferred time parameter is obtained by the requirement that the dynamics is expressed in terms of dimensionless quantities only \citep{BarbourKoslowskiMercati2013ProbOfTime}, but there is no reason why every value of that time parameter (which turns out to be equivalent to the ratio of the York parameter $T$ with some arbitrary reference value $T_0$) should indeed be obtained along the trajectory. Furthermore, shape dynamics differs from general relativity in its description of black holes \citep{Gomes2014,GomesHerczeg2014}, so the theory may escape the argument outlined in section \ref{mainreason}. In fact, in our opinion this view is the strongest reason \emph{not} to believe in the physicality of our proposed extension. Note though that shape dynamics is a different physical theory, which, while locally equivalent to general relativity in the CMC gauge, does differ in the global structures it allows and so is at least in principle\footnote{The observer would have to travel through the event horizon of a black hole in order to establish a difference.} observationally distinguishable. Hence, despite its similarities, it is not the focus of this paper. Meanwhile the view adopted here, ascribing a more fundamental role to time, remains viable at least until there is sufficient observational evidence to strongly favour theories (of quantum gravity, for example) which rely on a purely relational notion of time.

If we adopt the first point of view then a theoretical framework in which York time is fundamental imposes a restriction on inflationary potentials (or, more accurately, the complete matter content of the universe) and is therefore experimentally testable, provided cosmological flatness ($\Omega_k=0$) is conclusively established. At the time of writing the list of observationally unfalsfied inflationary potentials is still too long to make any definite conclusions in this context \citep{MartinEtAl2013} and there is no obvious correlation between potentials allowing for a smooth transition and those favoured by data on inflation. However, it will be interesting to see whether `York-time friendly' potentials are favoured or not as data improves. It is, however, important to bear in mind that some inflaton potentials are merely effective (derived from some underlying supersymmetry or string model, for example) and their form may not stay the same throughout the entire history of the universe. A more careful analysis of individual models is necessary.

At this stage, whether or not we should give credence to predictions concerning the unobservable epoch $T>0$ depends on the existence of other evidence for York time as a physically fundamental time parameter. For example, if dramatic progress is made in the development of some theory of quantum gravity that relies on York time, and observational evidence emerges that corroborates this theory, then this would be reason to take the extension seriously.

However, even in the absence of such evidence, at least when debating fundamental issues of time it is worth keeping the existence of such extensions in mind.

\paragraph{Acknowledgements.}
The author would like to thank Antony Valentini for helpful discussions and comments. This work was in part supported by the Foundational Questions Institute (fqxi.org).

%_________________________________________________________________________________
%BIBLIOGRAPHY

\bibliographystyle{abbrvnat}	%IF STYLE CHANGED, FIRST DELETE .AUX AND .BBL FILES BEFORE RECOMPILING
\bibliography{YorkTimeExtension_Bibloi}	%ENTER CORRECT RELATIVE PATH TO BIB-FILE

%\begin{thebibliography}{99}
%\small

%\end{thebibliography}

\end{document}